\documentclass[aps,prl,preprint,superscriptaddress,nobibnotes]{revtex4-2}
\usepackage{graphicx,color}
\usepackage{amsmath}
\usepackage{bm}
\usepackage{dcolumn}
\usepackage{ifthen}
\usepackage{multirow}
\usepackage{threeparttable}
\usepackage{siunitx}
\usepackage{hyperref}
\hypersetup{backref=true, pdfnewwindow=true, linkcolor=blue, colorlinks=true, anchorcolor=blue, citecolor=blue, filecolor=blue,  menucolor=blue, urlcolor=blue}
\usepackage{setspace}
\usepackage{lineno}

\begin{document}
\title{Atomic-Scale Detection of Néel Vector Switching in the Single-Layer A-type Antiferromagnet Cr$_2$S$_3$-2D}

\author{Affan Safeer}
\email{safeer@ph2.uni-koeln.de}
\affiliation{II. Physikalisches Institut, Universität zu Köln, Zülpicher Str. 77, D-50937 Köln, Germany}
\author{Calisa Dias}
\affiliation{Institute of Molecular Science, University of Valencia, Catedratico Jose Beltrán 2, 46980 Paterna, Spain}
\author{Mahdi Ghorbani-Asl}
\affiliation{Institute of Ion Beam Physics and Materials Research, Helmholtz-Zentrum Dresden-Rossendorf, D-01328 Dresden, Germany}
\author{Abdallah Karaka}
\affiliation{II. Physikalisches Institut, Universität zu Köln, Zülpicher Str. 77, D-50937 Köln, Germany}
\author{Pradyumna Bawankule}
\affiliation{Institute of Molecular Science, University of Valencia, Catedratico Jose Beltrán 2, 46980 Paterna, Spain}
\author{Weibin Li}
\affiliation{ALBA Synchrotron Light Source, Cerdanyola del Vallès, 08290 Barcelona, Spain}
\author{Pierluigi Gargiani}
\affiliation{ALBA Synchrotron Light Source, Cerdanyola del Vallès, 08290 Barcelona, Spain}
\author{Wouter Jolie}
\affiliation{II. Physikalisches Institut, Universität zu Köln, Zülpicher Str. 77, D-50937 Köln, Germany}
\author{Arkady V. Krasheninnikov}
\affiliation{Institute of Ion Beam Physics and Materials Research, Helmholtz-Zentrum Dresden-Rossendorf, D-01328 Dresden, Germany}
\author{Amilcar Bedoya-Pinto}
\affiliation{Institute of Molecular Science, University of Valencia, Catedratico Jose Beltrán 2, 46980 Paterna, Spain}
\author{Thomas Michely}
\affiliation{II. Physikalisches Institut, Universität zu Köln, Zülpicher Str. 77, D-50937 Köln, Germany}
\author{Jeison Fischer}
\email{jfischer@ph2.uni-koeln.de}
\affiliation{II. Physikalisches Institut, Universität zu Köln, Zülpicher Str. 77, D-50937 Köln, Germany}

\begin{abstract}
The detection of Néel vector switching in a single-layer A-type antiferromagnet marks an important step toward functional two-dimensional spintronics. Here, Cr$_2$S$_3$-2D, grown on graphene on Ir(110), is established as a first single-layer A-type antiferromagnet. Spin-polarized scanning tunneling microscopy reveals hysteresis loops with a large switching field and a pronounced dependence on island size. X-ray magnetic circular dichroism at the Cr L$_{2,3}$ edges exhibits a tiny signal with a linear magnetic field dependence, consistent with a nearly compensated antiferromagnetic ground state and a Néel temperature of about 160\,K. Quantitative analysis of the island-size dependence of the switching field, together with first principles calculations, indicates a slight imbalance between the magnetic moments of the two Cr planes of Cr$_2$S$_3$-2D when supported on a substrate. This imbalance results in a net magnetization for the A-type antiferromagnet, which enables the 180$^\circ$ rotation of the Néel vector. Moreover, Cr$_2$S$_3$-2D retains its magnetic properties after several days of exposure to air.
\end{abstract}

\maketitle
\newpage

\section{Introduction}

\begin{figure}[htbp]
\centering
\includegraphics[width=0.95\textwidth]{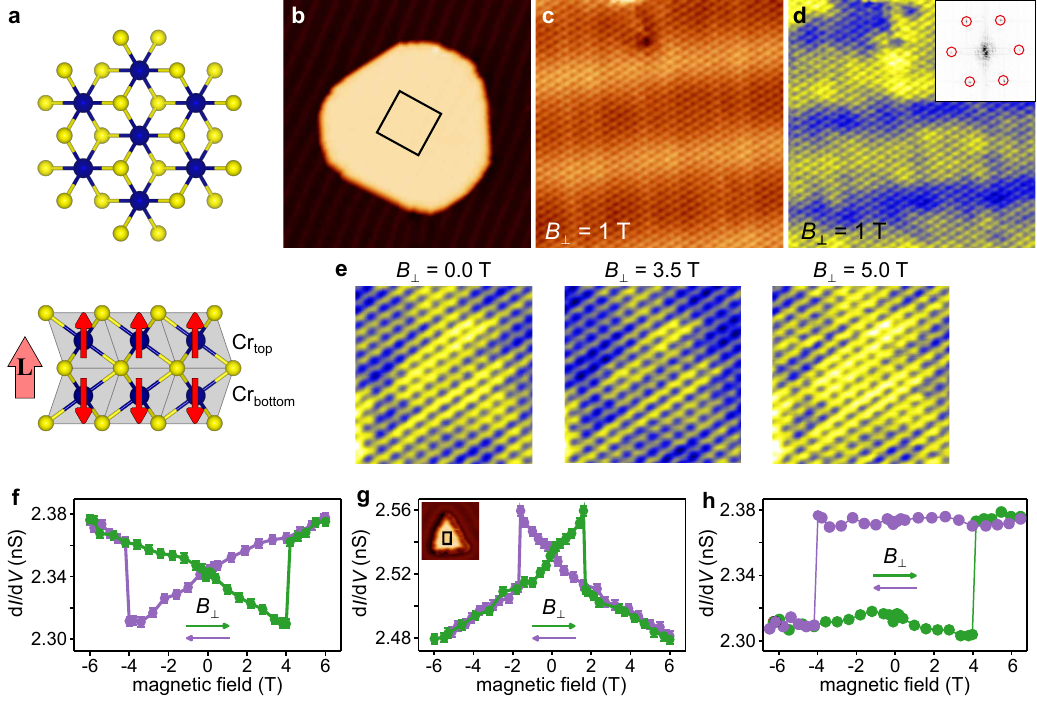}
\caption{SP-STM of single-layer Cr$_2$S$_3$-2D. \textbf{a}, Top and side view ball models of single-layer Cr$_2$S$_3$-2D with the two Cr planes labeled Cr$_\mathrm{top}$ and Cr$_\mathrm{bottom}$. Spins (red arrows) are attached to the Cr atoms in the side view symbolizing A-type AF order. The Néel vector (\textbf{L}) is indicated. Yellow balls: S; blue balls: Cr. \textbf{b}, STM overview topograph ($V_\mathrm{b} = 1$\,V, $I_\mathrm{t} = 20$\,pA) of a Cr$_2$S$_3$-2D island on Gr/Ir(110). \textbf{c}, Atomically resolved SP-STM topograph ($V_\mathrm{b} = 100$\,mV, $I_\mathrm{t} = 1$\,nA) of Cr$_2$S$_3$-2D recorded with an out-of-plane field of $B_\perp = 1$\,T. Spin-polarized tips were calibrated on a reference sample (Figure~S1, Supporting Information). \textbf{d}, $\mathrm{d}I/\mathrm{d}V$ map of Cr$_2$S$_3$-2D recorded simultaneously with (\textbf{c}). The inset shows the fast Fourier transform of the $\mathrm{d}I/\mathrm{d}V$ map. \textbf{e}, $\mathrm{d}I/\mathrm{d}V$ maps ($V_\mathrm{b} = 100$\,mV, $I_\mathrm{t} = 1$\,nA) recorded at out-of-plane fields as indicated. \textbf{f}, Average $\mathrm{d}I/\mathrm{d}V$ value as a function of the magnetic field. Each data point is the average over a $\mathrm{d}I/\mathrm{d}V$ map as in (\textbf{e}). All maps are taken from exactly the same area. \textbf{g}, Average $\mathrm{d}I/\mathrm{d}V$ extracted from $\mathrm{d}I/\mathrm{d}V$ maps taken on a Co island on Cu(111) as a function of magnetic field. The inset shows the $\mathrm{d}I/\mathrm{d}V$ map ($V_\mathrm{b} = -300$\,mV, $I_\mathrm{t} = 1$\,nA) of the Co island on Cu(111), black rectangle shows where data is acquired. \textbf{h}, De-convoluted field-dependent $\mathrm{d}I/\mathrm{d}V$ response of the Cr$_2$S$_3$-2D, achieved by subtracting the tip contribution established in (\textbf{g}) from (\textbf{f}). Image size: (\textbf{b}) 50\,nm $\times$ 50\,nm, (\textbf{c},\textbf{d}) 10\,nm $\times$ 10\,nm, (\textbf{e}) each image 4\,nm $\times$ 4\,nm, and inset of (\textbf{g}) 15\,nm $\times$ 15\,nm.
}
\label{fig_spstm}
\end{figure}

Antiferromagnets (AFs) are emerging as key building blocks for spintronics because they produce no stray field, are robust against external perturbations, and allow ultrafast spin dynamics with the possibility of efficient electrical and optical control \cite{Baltz2018, Zelezny2018, Guo2025}. An important aspect of spintronics with AFs is the ability to manipulate and switch the Néel vector, which enables active control of antiferromagnetic order \cite{Han2024, Wang2026}. Extending these advantages to two-dimensional (2D) materials is particularly appealing, as this promises ultrathin, high-density spintronic elements and new regimes of low-dimensional magnetism \cite{Liu2020, Zhao2025, Jia2025}. In single-layer AFs zigzag and stripy orders have been identified \cite{Yang2023,Xian2022,Ni2021}, but A-type antiferromagnetic order (Figure~\ref{fig_spstm}a) -- where ferromagnetic spin layers are antiferromagnetically coupled, providing a layer-resolved Néel vector for more direct control -- has so far only been observed in bilayer van der Waals systems, such as CrI$_3$ \cite{Huang2017} and CrSBr \cite{Lee2021}. 

To realize a true single-layer A-type AF, we consider the versatile material class of chromium chalcogenides. For example, metastable 1$T$-CrTe$_2$ has emerged as van der Waals crystal with room-temperature ferromagnetism \cite{Freitas2015}, whereas in the single-layer limit it is a zigzag AF with Néel temperature $T_\mathrm{N}$ $\approx$ 140\,K \cite{Xian2022, Kushwaha2025}. In thermodynamic equilibrium CrTe$_3$ is the only chromium chalcogenide van der Waals crystal \cite{Klepp1982, Chattopadhyay94}. The overwhelming majority of chromium chalcogenide compounds are covalently bonded \cite{Jellinek1957,Chattopadhyay94,Wehmeier70}.  Therefore, it is not surprising that in recent years several chromium chalcogenide compounds have been synthesized as fully covalently bonded ultrathin films or even single unit cell thickness 2D materials \cite{Chuang2025, Song2024, Kushwaha2025, safeer2025}. 

Here we focus on one of these covalently bonded 2D materials of single unit cell thickness, namely Cr$_2$S$_3$-2D, whose structure was recently described \cite{safeer2025} and which does not exist in bulk. It can be pictured as five vertically stacked hexagonal planes in a sequence of S--Cr--S--Cr--S as shown in Figure~\ref{fig_spstm}a. Each Cr$^{3+}$ ion in Cr$_2$S$_3$-2D is octahedrally coordinated by six S ions with average charge state $2-$. Since Cr$_2$S$_3$-2D is different in structure from bulk Cr$_2$S$_3$, to avoid confusion, we append "-2D" to its chemical formula. 

Density functional theory (DFT) calculations for Cr$_2$S$_3$-2D invariably predict ferromagnetic ordering within each Cr plane \cite{Li2023,Zhang2021,Chen25,safeer2025} dominated by in-plane superexchange with S--Cr--S bond angles close to 90$^\circ$. For coupling between the planes, the picture is more complex, and both direct-exchange favoring antiferromagnetic coupling and super-exchange Cr--S--Cr paths favoring ferromagnetic or antiferromagnetic coupling are present \cite{Zhang2021}. This ambiguity is reflected in DFT calculations, where the interplane coupling changes from antiferromagnetic to ferromagnetic when the Hubbard $U$ parameter exceeds $\approx 2$\,eV \cite{Zhang2021,safeer2025}. Thus, based on DFT, Cr$_2$S$_3$ can be assumed to be an A-type AF (small $U$) or a ferromagnet (large $U$). The latter is assumed by most authors \cite{Li2023,Zhang2021,Chen25}.

Recently, two A-type AFs of the NiAs-type crystal structure family have been identified as altermagnets \cite{Liber2022,Guo2025}, namely CrSb \cite{Reimers24} and MnTe \cite{Osumi24}. If Cr$_2$S$_3$-2D, which is the ultimately thin NiAs-type 2D material, could be established as A-type AF, exploring low-dimensional altermagnetism and its associated transport phenomena would become accessible. 

Here, by combining spin-polarized scanning tunneling microscopy (SP-STM) with X-ray magnetic circular dichroism (XMCD) measurements on Cr$_2$S$_3$-2D, we encounter an apparent paradox. While seemingly ferromagnetic hysteresis loops with switching fields of several Tesla are observed by SP-STM, XMCD proves a negligible magnetic moment of the Cr atoms implying antiferromagnetic order. The puzzle is solved by realizing that switching characterized by a 180$^\circ$ rotation of the Néel vector is enabled by a small uncompensated moment which arises from the interaction with the substrate.

\section{Results and discussion}

\subsection{Hysteresis loops on Cr$_2$S$_3$-2D islands}

To investigate the magnetic properties of Cr$_2$S$_3$-2D, we obtained at 4.2\,K atomically and spin-resolved STM images together with $\mathrm{d}I/\mathrm{d}V$ maps of a Cr$_2$S$_3$-2D island, as shown in Figs.~\ref{fig_spstm}b-e.  Both the SP-STM topograph (Figure~\ref{fig_spstm}c) and the corresponding $\mathrm{d}I/\mathrm{d}V$ map (Figure~\ref{fig_spstm}d) obtained at out-of-plane field of 1~T reveal the hexagonal lattice of Cr$_2$S$_3$-2D with a lattice constant of 0.340\,nm, superimposed on the moiré pattern of the underlying Gr/Ir(110) substrate (visible as horizontal bright lines in Figure~\ref{fig_spstm}c,d). Although at the bias of $V_\mathrm{b} = 100$\,mV a magnetic field dependent change in the $\mathrm{d}I/\mathrm{d}V$ signal is observed (see Figure~S2, Supporting Information), no additional superstructure or magnetic superlattice is detected, as confirmed by the fast Fourier transform of the $\mathrm{d}I/\mathrm{d}V$ map shown in the inset of Figure~\ref{fig_spstm}d.

However, when the magnetic field is varied, the overall signal of the $\mathrm{d}I/\mathrm{d}V$ maps displays a slight change in contrast (Figure~\ref{fig_spstm}e). To systematically probe this behavior, $\mathrm{d}I/\mathrm{d}V$ maps are acquired over a magnetic field ranging between $+6$\,T and $-6$\,T. To obtain a sufficient signal-to-noise ratio, their $\mathrm{d}I/\mathrm{d}V$ value is averaged in an area of 16\,nm$^2$ and plotted as a function of magnetic field in Figure~\ref{fig_spstm}f. The field-dependent $\mathrm{d}I/\mathrm{d}V$ values show a butterfly-shaped hysteresis, with a quasi-linear response and symmetrical abrupt jumps at $\pm4$\,T.

In SP-STM the $\mathrm{d}I/\mathrm{d}V$ signal depends on the spin polarization of both the STM tip and the sample's top plane~\cite{Wortmann2001}. To isolate the tip contribution from the observed hysteresis behavior, after acquiring the field dependent $\mathrm{d}I/\mathrm{d}V$ signal on Cr$_2$S$_3$-2D, the same tip is used to measure a Co island on Cu(111) (inset of Figure~\ref{fig_spstm}g) which is a well-characterized ferromagnet with out-of-plane easy axis~\cite{Ouazi2012}. The resulting curve exhibits a similar, but inverted butterfly-shaped hysteresis with abrupt jumps at $\pm1.6$~T (Figure~\ref{fig_spstm}g), corresponding to the magnetization switching of the Co island. The gradual change originates from the field-dependent spin polarization of the STM tip~\cite{Ouazi2012, Rodary2008}. The butterfly-shape is inverted due to the negative spin polarization of the Co islands at the measurement bias~\cite{Oka2010}.

After removing the tip contribution by subtracting the slope extracted from the Co island hysteresis, the resulting field-dependent $\mathrm{d}I/\mathrm{d}V$ curve exhibits a square hysteresis (Figure~\ref{fig_spstm}h). This result is consistent with long-range ferromagnetic ordering within the top Cr$_\mathrm{top}$ plane, and the abrupt change of $\mathrm{d}I/\mathrm{d}V$ (in Figure~\ref{fig_spstm}f) signifies its reversal. The square shape obtained in our analysis indicates a dominant out-of-plane easy-axis. This is further corroborated by an unchanged $\mathrm{d}I/\mathrm{d}V$ signal of the in-plane field magnetization curve shown in Figure~S3 (Supporting Information).

The switching field is exceptionally large and unattained in previously reported 2D ferromagnetic materials. For a ferromagnet, such a large switching field is indicative of strong magnetic anisotropy. However, if the ferromagnetic top Cr$_\mathrm{top}$ plane would be antiferromagnetically coupled to the bottom Cr$_\mathrm{bottom}$ plane, the large switching field could reflect a small residual net magnetic moment to which the external field couples only weakly.

\subsection{A-type antiferromagnetic ground state in Cr$_2$S$_3$-2D}

\begin{figure*}[!ht]
\centering
\includegraphics[width=\textwidth]{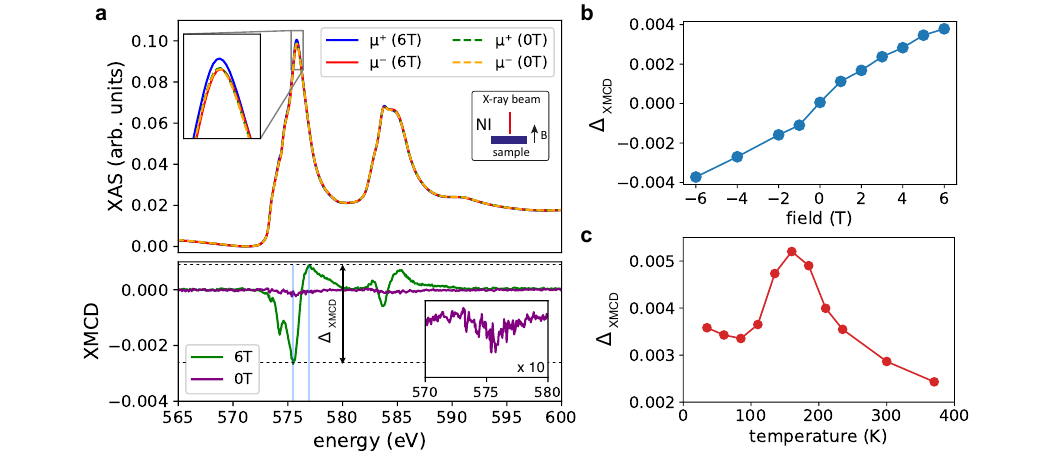}
\caption{XMCD of single-layer Cr$_2$S$_3$-2D at normal incidence. \textbf{a}, Upper panel: Cr L$_{2,3}$ edges XAS of single-layer Cr$_2$S$_3$-2D on Gr/Ir(110) acquired in total electron yield mode for positive ($\mu^+$) and negative ($\mu^-$) helicities at 3.5\,K with 0\,T and 6\,T. Insets on the left and right show the zoom of the Cr L$_3$ edge and the measurement schematics, respectively. Lower panel: Resulting XMCD signal ($\mu^--\mu^+$). Inset shows 0\,T XMCD signal multiplied by a factor of 10. \textbf{b}, Peak-to-peak span of XMCD Cr L$_3$ edge ($\Delta_\mathrm{XMCD}$) as a function of magnetic field at 3.5\,K. \textbf{c}, Temperature-dependent $\Delta_\mathrm{XMCD}$ of Cr L$_3$ edge with 6\,T applied magnetic field. The $\Delta_\mathrm{XMCD}$ values are extracted in both field- and temperature-dependent variation from the values in photon energy marked by lines in the XMCD panel of (\textbf{a}), corresponding to the minimum and maximum signal at the Cr L$_3$ edge, and taking the difference ($\Delta$) between them.
}
\label{Fig_XMCD}
\end{figure*}

To determine the actual magnetic state, we performed XMCD measurements on single-layer Cr$_2$S$_3$-2D. XMCD is able to probe both top and bottom Cr planes with high sensitivity in total electron yield, which in addition, allows element-specific detection. The XMCD is obtained as the difference between the left ($\mu$$^-$) and the right ($\mu$$^+$) circularly polarized X-ray absorption spectroscopy (XAS) at the Cr L$_{2,3}$ edges in normal incidence (NI, $\phi$ = 0$^\circ$) (Figure~\ref{Fig_XMCD}) and grazing incidence (GI, $\phi$ = 70$^\circ$) (Figure~S4, Supporting Information) with field parallel to the incident beam. These geometries are sensitive to out-of-plane and in-plane spin components, respectively. The XAS exhibits the Cr L$_{3}$ and Cr L$_{2}$ peaks at 575.6\,eV and 583.2\,eV with an overall shape consistent with the XAS of bulk chromium sulfide~\cite{Yaji2004}.

The XMCD spectrum in NI (Figure~\ref{Fig_XMCD}a) exhibits a small signal at 6\,T. Assuming the validity of XMCD sum rules \cite{Chen_PRL95,Bedoya-Pinto2021}, the estimated magnetic moment per Cr atom is $\leq 0.1\,\mu_{\mathrm{B}}$ at 6\,T. This small XMCD signal is consistent with field-induced spin canting of an antiferromagnetic system. Moreover, the nearly linear dependence of the XMCD intensity at the L$_3$ edge $\Delta_\mathrm{XMCD}$ on the applied magnetic field (Figure~\ref{Fig_XMCD}b) supports an antiferromagnetic ground state, in which the external field enhances the spin canting.

At zero field, the XMCD signal almost vanishes (Figure~\ref{Fig_XMCD}a), but not completely. Such a residual XMCD signal could indicate uncompensated, magnetically correlated spins in Cr$_2$S$_3$-2D. A remanent magnetic signal of all Cr atoms would imply a net magnetic moment per atom $\lesssim 5 \times 10^{-3}\,\mu_{\mathrm{B}}$ at 0\,T.

To further clarify the nature of magnetic coupling, we performed temperature-dependent XMCD measurements for $B = 6$\,T (Figure~S5, Supporting Information). In Figure~\ref{Fig_XMCD}c, we represent $\Delta_\mathrm{XMCD}$ showing a non-monotonic behavior with respect to temperature, reaching a maximum around 160\,K above which it decreases steadily.

These results are interpreted as follows: At $T<T_\mathrm{N}$, strong interplane exchange coupling enforces antiparallel alignment between the Cr planes, thermal excitation progressively weakens this coupling and enables the field-induced spin canting more easily, resulting in an increase of the XMCD signal. As the temperature approaches $T_\mathrm{N}$, spin fluctuations and spin flip give rise to a maximum XMCD signal, beyond which the signal decreases as the system becomes paramagnetic. The observed non-monotonic behavior thus reflects a typical temperature-dependent susceptibility of an AF, with a transition to a paramagnetic phase at a Néel temperature $T_\mathrm{N} \approx 160$\,K. This Néel temperature is also identified by X-ray magnetic linear dichroism signal, which drops to zero at $T_\mathrm{N}$ (Figure~S6, Supporting Information).

The combination of the ferromagnetic response of the topmost Cr plane observed by SP-STM (Figure~\ref{fig_spstm}) and the compensated antiferromagnetic structure revealed by XMCD (Figure~\ref{Fig_XMCD}) demonstrates that Cr$_2$S$_3$-2D is an A-type AF.

Although antiferromagnetic order would ideally lead to zero net magnetization, switching as observed in Figure~\ref{fig_spstm} reveals that a magnetization reversal mechanism is available, that is, a rotation of the Néel vector by $180^\circ$. Switching may arise from uncompensated magnetic moments, typically associated with surface, interface, or edge atoms, which couple to the applied field and induce rotation of the Néel vector. To elucidate the origin of the SP-STM hysteresis, we carry out measurements for different island sizes.

\subsection{Size dependent Cr$_2$S$_3$-2D island switching}

\begin{figure*}[!ht]
\centering
\includegraphics[width=0.9\textwidth]{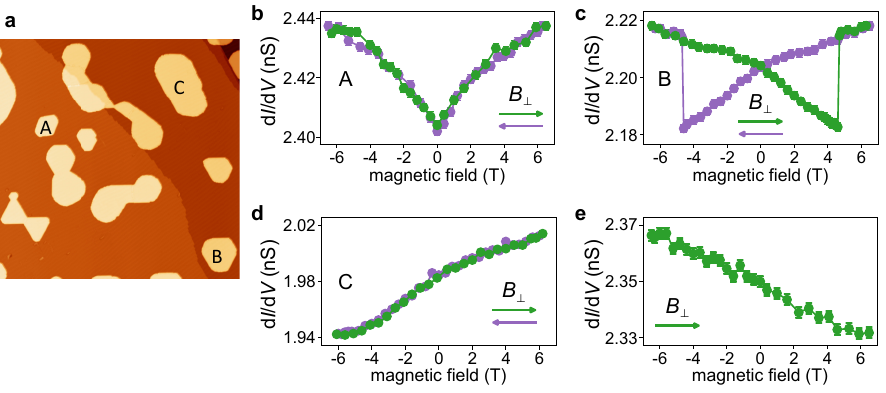}
\caption{Size-dependent Cr$_2$S$_3$-2D island switching. \textbf{a}, STM overview topograph ($V_\mathrm{b} = 1$\,V, $I_\mathrm{t} = 20$\,pA, 200~nm $\times$ 200~nm) with Cr$_2$S$_3$-2D islands. \textbf{b}-\textbf{d}, Average $\mathrm{d}I/\mathrm{d}V$ as a function of field for islands A, B, and C in (\textbf{a}), respectively. \textbf{e}, Average $\mathrm{d}I/\mathrm{d}V$ of an island of similar size as C, but taken in a different location.
}
\label{fig_size}
\end{figure*}

Representative field-dependent $\mathrm{d}I/\mathrm{d}V$ curves for small, intermediate, and large islands labeled A, B, and C in the STM overview of Figure~\ref{fig_size}a are shown in Figs.~\ref{fig_size}b-d. In addition to the butterfly-shaped $\mathrm{d}I/\mathrm{d}V (B_\perp)$ curve (Figure~\ref{fig_size}c) already discussed for Figure~\ref{fig_spstm}f, the small island A exhibits a V-shaped curve (Figure~\ref{fig_size}b), while large islands show curves without any switching with positive (Figure~\ref{fig_size}d) or negative (Figure~\ref{fig_size}e) slope.

The interpretation of the new types of magnetization curves is as follows. When the switching field is zero (the topmost plane behaves like a superparamagnet) or close to zero, the hysteretic jump in $\mathrm{d}I/\mathrm{d}V (B_\perp)$ takes place at zero or close to zero field, that is, it is absent as in Figure~\ref{fig_size}b. $\mathrm{d}I/\mathrm{d}V (B_\perp)$ curves without switching and positive or negative slope of $\mathrm{d}I/\mathrm{d}V (B_\perp)$ are due to islands where the switching field is not within the range of measurement, i.e., the magnetization of the island is fixed (Figs.~\ref{fig_size}d and \ref{fig_size}e). The different signs of slope arise from the two possible topmost Cr plane magnetizations.

The statistical analysis of over 25 islands (Figure~\ref{fig_DeltaE}a) identifies a monotonic increase of the switching field ($\mu_\mathrm{0}H_\mathrm{sw}$) with island size given by its number of Cr atoms ($N_\mathrm{Cr}$). Islands smaller than $N_\mathrm{Cr} \approx$ 6000~atoms (300~nm$^2$) switch near or at zero field, islands between $\approx$ 6000 and $\approx$ 30000~Cr atoms exhibit finite switching field, and islands above $N_\mathrm{Cr} > 30000$~atoms (1500~nm$^2$) show no switching within the measured magnetic field range of $\pm6.5$~T. 

\subsection{The mechanism of Néel vector switching}

To gain further insight into the reversal mechanism in Cr$_2$S$_3$-2D islands, we extract the energy barrier $\Delta E$ from the measured switching fields $\mu_\mathrm{0}H_{\mathrm{sw}}$. In an antiferromagnetic particle, the dominant contribution to $\Delta E$ for a coherent rotation of the Néel vector $\bf{L=m_\mathrm{Cr_{top}}-m_\mathrm{Cr_{bottom}}}$ originates from the anisotropy of all Cr atoms, while the applied field couples to a small uncompensated magnetic moment $\mu_u$. Here $\bf{m}_\mathrm{Cr_{top}}$ and $\bf{m}_\mathrm{Cr_{bottom}}$ are the total magnetic moments of the two Cr sublattices.

\begin{figure*}[!ht]
\centering
\includegraphics[width=1.0\textwidth]{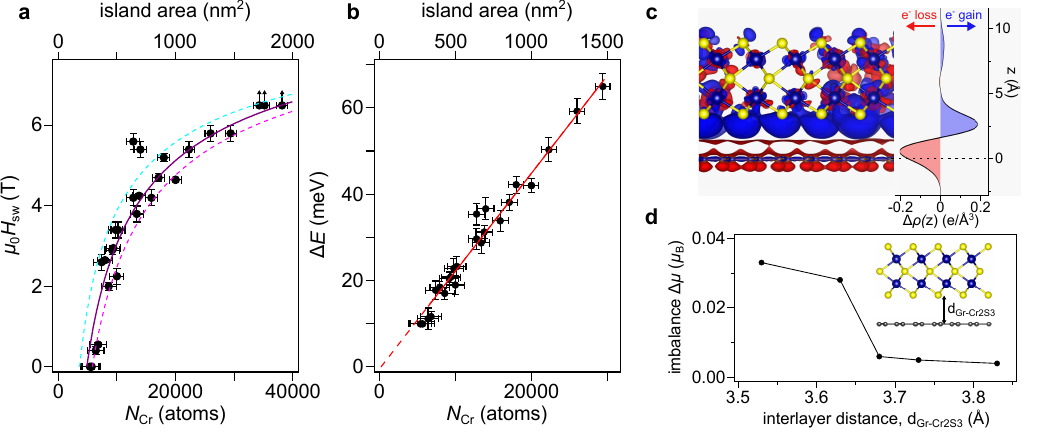}
\caption{Quantitative estimation of uncompensated moments and charge transfer from substrate. \textbf{a}, Switching field $\mu_\mathrm{0}H_{\mathrm{sw}}$ and \textbf{b}, energy barrier ($\Delta E$) of islands as a function of island size. $\Delta E$ is extracted from $\mu_\mathrm{0}H_{\mathrm{sw}}$ in (\textbf{a}) by using equation~\ref{eq:Hsw}. Violet and red lines in (\textbf{a}) and (\textbf{b}) are fits based on equation~\ref{eq:Hsw} assuming the uncompensated magnetic moment $\Delta \mu = 8 \times 10^{-3}\,\mu_{\mathrm{B}}$ per unit cell to stem from the difference in the Cr atomic moments in the top Cr$_\mathrm{top}$ and bottom Cr$_\mathrm{bottom}$ planes. Magenta and cyan dashed lines are obtained with smaller $\Delta \mu = 6 \times 10^{-3}\,\mu_{\mathrm{B}}$ and larger $\Delta \mu = 1 \times 10^{-2}\,\mu_{\mathrm{B}}$. \textbf{c}, Charge density difference plot overlaid with ball model of the relaxed DFT geometries for Cr$_2$S$_3$-2D/Gr. Sphere colours: S: yellow; Cr: blue; C: grey. Charge gain (loss) represented by blue (red). \textbf{d}, $\Delta \mu$ as a function of distance between Gr and Cr$_2$S$_3$-2D, $d_\mathrm{Gr-Cr_2S_3}$ as indicated in the inset.}
\label{fig_DeltaE}
\end{figure*}

Thermal fluctuations assist the system in overcoming $\Delta E$; evidence for this statement is that lowering the temperature from 4.2~K to 1.7~K increases the switching field (Figure~S7, Supporting Information). For the case of an uniaxial anisotropy and magnetization reversal, that is, switching of the Néel vector $\bf{L}$ by $180^\circ$ at a temperature $T$, the relation between the $\mu_\mathrm{0}H_{\mathrm{sw}}$ and $\Delta E$ for a measurement time $t_{\mathrm{meas}}$ is given by a Sharrock-type expression \cite{Sharrock94},
\begin{equation}
\mu_\mathrm{0} H_{\mathrm{sw}} = \frac{2\Delta E}{\mu_u}\left[1 - \left(\frac{k_B T}{\Delta E}\ln\!\left(\frac{t_{\mathrm{meas}}}{\tau_0}\right)\right)^{1/2}\right],
\label{eq:Hsw}
\end{equation}
where $k_\mathrm{B}$ is the Boltzmann constant and $\tau_0$ is the attempt time. In our analysis, we use a typical value of $\tau_0 = 10^{-10}\,\mathrm{s}$ for the attempt time \cite{Wernsdorfer01} and $t_{\mathrm{meas}} = 100\,\mathrm{s}$, corresponding to the duration of the measurement. 

To model the total uncompensated magnetic moment $\mu_u$ in equation (1), we assume a difference $\Delta \mu = \mu_\mathrm{Cr_{top}} -\mu_\mathrm{Cr_{bottom}}$ in the atomic magnetic moments of the two Cr atom planes leading to $\mu_u = 1/2 N_{\mathrm{Cr}} \Delta \mu$ with $N_{\mathrm{Cr}}$ being the total number of Cr atoms in an island ($1/2 N_{\mathrm{Cr}}$ in each Cr plane). Assuming a single value $\Delta \mu$, equation (1) is solved for each island size characterized by $N_{\mathrm{Cr}}$. Figure~\ref{fig_DeltaE}b shows the energy barrier $\Delta E(N_{\mathrm{Cr}})$ for $\Delta \mu = 8 \times 10^{-3}\,\mu_{\mathrm{B}}$. The data points lie along a linear $\Delta E$ dependence on $N_{\mathrm{Cr}}$. This behavior is consistent with an effective Stoner--Wohlfarth-type description of the antiferromagnetic energy barrier, in which the energy barrier is given by $\Delta E = K N_{\mathrm{Cr}}$, where $K$ is the effective anisotropy energy per Cr atom. The red line in Figure~\ref{fig_DeltaE}b is a linear fit through the data points with the slope being $K=2.3~\mu$\,eV/Cr~atom. Replacing in equation (1) $\Delta E$ by $K N_{\mathrm{Cr}}$ and using $\mu_u = 1/2 N_{\mathrm{Cr}} \Delta \mu$ we predict the functional dependence $H_{\mathrm{sw}}(N_{\mathrm{Cr}})$ shown as violet line in Figure~\ref{fig_DeltaE}a. It is evident that the choice of $\Delta \mu = 8 \times 10^{-3}\,\mu_{\mathrm{B}}$ provides a decent fit to the experimental data. Smaller ($\Delta \mu = 6 \times 10^{-3}\,\mu_{\mathrm{B}}$, magenta dashed line) and larger ($\Delta \mu = 1 \times 10^{-2}\,\mu_{\mathrm{B}}$, cyan dashed line) values of $\Delta \mu$ provide inferior fits. We emphasize that $\Delta \mu$ is the only fit parameter in the entire procedure defining $\mu_u, \Delta E, K$ and $H_{\mathrm{sw}}(N_{\mathrm{Cr}})$. Recalculating $\Delta \mu$ into a fictitious effective magnetic moment per Cr atom yields a value of $4 \times 10^{-3}\,\mu_{\mathrm{B}}$ per atom which is consistent with our rough estimate from XMCD at 0\,T.
As an alternative model for $\mu_u$ in equation~\ref{eq:Hsw}, we tested the assumption that it stems from the uncompensated spins of the Cr island edge atoms and assigned to each edge atom a magnetic moment $\mu_e$. As documented by Figure~S8 (Supporting Information), this model is clearly worse than the one of Figure~\ref{fig_DeltaE}. Edge effects are not dominant for switching of islands.

Interface-induced magnetic effects are well established and include phenomena such as proximity-induced magnetization, spin-orbit coupling driven surface and interface effects, and magnetism arising from charge transfer \cite{Hellman17}. The latter is well known for oxide superlattices \cite{Keunecke20,Bange22}. In 2D heterostructures, charge transfer has also been shown to modify interplane exchange interactions \cite{Hong2025}. For Cr$_2$S$_3$-2D charge transfer from the substrate is inferred from the fact that ab initio calculations consistently show freestanding Cr$_2$S$_3$ to be semiconducting \cite{Zhang2021,Li2023,Chen25,safeer2025}, whereas when grown on Gr/Ir(110) it is metallic \cite{safeer2025}. The electron transfer from the substrate into the Cr$_2$S$_3$-2D conduction band is accompanied by the formation of an interface dipole, which originates from vacuum level alignment \cite{safeer2025}.

The charge transfer from Gr to Cr$_2$S$_3$-2D and the resulting imbalance in the magnetic moments are captured by DFT calculations. We employed a DFT supercell (Figure~S9, Supporting Information) of a $5\times5$ Cr$_2$S$_3$-2D layer {matching} a $7\times7$ Gr layer almost free of strain. Upon structural relaxation, the interlayer distance converges to $d_\mathrm{Gr-Cr_2S_3}=3.53$~\AA. At this separation, charge transfer occurs from Gr to Cr$_2$S$_3$-2D, leading to charge accumulation at the bottom S plane, and a corresponding depletion on the Gr side, as illustrated in Figure~\ref{fig_DeltaE}c. Without Gr (i.e., freestanding Cr$_2$S$_3$-2D), we find an average magnetic moment of 2.79 $\mu_\mathrm{B}$ per Cr atom in each Cr plane. However, with Gr underneath, because the interfacial charge redistribution modifies the occupation of the spin-polarized states in Cr$_2$S$_3$-2D, the average magnetic moment in the bottom Cr$_\mathrm{bottom}$ plane decreases to 2.77 $\mu_\mathrm{B}$ per Cr atom, while the top Cr$_\mathrm{top}$ plane exhibits a magnetic moment of 2.80 $\mu_\mathrm{B}$ (Figure~S7, Supporting Information). The imbalance $\Delta \mu = 0.03 \mu_\mathrm{B}$ is of the same order of magnitude as the rough estimate based on our model. Increasing the distance between Gr and Cr$_2$S$_3$-2D reduces $\Delta \mu$, restoring the magnetic moments to values close to those of freestanding Cr$_2$S$_3$-2D.

Taking a closer look at the fit curve in Figure~\ref{fig_DeltaE}a, we find that not only the size dependence of $\mu_\mathrm{0}H_{\mathrm{sw}}$ but also the onset of switching at a finite island size is well reproduced. Small islands are not in a blocked state, because $T>\frac{KN}{25 k_\mathrm{B}}$ \cite{Bean1959}. Less certain is the description of the transition from switching to non-switching. The three data points of the largest islands, that did not switch at 6.5\,T (recognizable by the up-arrows attached to them), lie either on or even above the fit curve, although their switching field could be much higher. 

Another uncertainty is the fact that switching of small islands was found to be slightly affected by the tunneling set point during spectroscopy (Figure~S10, Supporting Information). Variation of the tunneling set point changes the tip electric field, the Joule heating, or even spin transfer torque. Irrespective of the precise origin, the effect on switching for small islands may have influenced our fit and could have led to an underestimation of $\mu_\mathrm{0}H_{\mathrm{sw}}$ for large islands.

Our analysis suggests that the Cr$_2$S$_3$-2D islands can be regarded as an effective ferrimagnetic system, in which the top and bottom Cr plane sublattices have a slight difference in magnetic moment, actively driving the reversal process. While responding to the field via the Zeeman term, the uncompensated moments allow a field-driven rotation of the Néel vector $\bf{L}$ by $180^\circ$. Because SP-STM is a surface-sensitive technique, an inversion of the $\bf{L}$ in this A-type AF is readily detected through the change of magnetization in the top Cr$_\mathrm{top}$ plane as in Figure~\ref{fig_spstm} and in Figure~\ref{fig_size}. This picture provides a first-order description that captures the key features of thermally activated Néel vector reversal.


\subsection{Magnetic order in Cr$_2$S$_3$-2D is air-stable}

\begin{figure*}[!ht]
\centering
\includegraphics[width=0.5\textwidth]{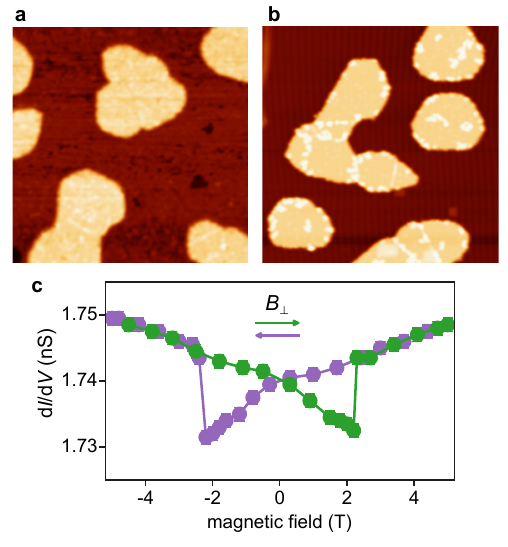}
\caption{Air stability of single-layer Cr$_2$S$_3$-2D. \textbf{a}, STM topograph of Cr$_2$S$_3$-2D on Gr/Ir(110) taken after exposure to air for two days and subsequent reintroduction to UHV without any additional treatment. \textbf{b}, STM topograph of the same sample after annealing at 600\,K for 60\,s. \textbf{c}, Averaged $\mathrm{d}I/\mathrm{d}V$ signal as a function of magnetic field.}
\label{fig_air}
\end{figure*}

To examine the robustness of the new 2D antiferromagnet, we exposed the sample to ambient conditions for 2 days. The STM image of Figure~\ref{fig_air}a acquired after reintroducing the sample to ultrahigh vacuum (UHV) without any additional treatment shows intact Cr$_2$S$_3$-2D islands with preserved morphology. However, adsorbates are present on both graphene and Cr$_2$S$_3$-2D, impairing SP-STM measurements. To remove these adsorbates, the sample is briefly annealed to 600\,K in UHV. As shown in Figure~\ref{fig_air}b, while adsorbates remain at edges and grain boundaries, they desorb from the island area and leave its surface largely clean. Subsequent $\mathrm{d}I/\mathrm{d}V (B_\perp)$ measurements (Figure~\ref{fig_air}c) reproduce the same characteristic hysteresis observed prior to air exposure. The persistence of the hysteresis confirms that the magnetic ground state of single-layer Cr$_2$S$_3$-2D remains unchanged after prolonged exposure to ambient conditions. 

\section{Conclusions}

In summary, small Cr$_2$S$_3$-2D islands display unblocked behavior in SP-STM at 4.2\,K, above an island size of $6 \times 10^3$ Cr atoms exhibit a clear hysteresis, and switching ceases for islands with a size above $3 \times 10^4$ Cr atoms within the experimental limit of 6.5\,T applied. XMCD at 3.5\,K and the Cr L$_{2,3}$ edges exhibit only a small signal for 6\,T, of which only a small fraction is left at remanence (0\,T). The overall XMCD response shows a linear dependence on the external magnetic field and a non-monotonic temperature dependence, aligning with nearly compensated antiferromagnetic order and an ordering temperature of about 160\,K. Exposure of Cr$_2$S$_3$-2D to the ambient does not change its magnetic response. Thus, Cr$_2$S$_3$-2D is an air stable single-layer A-type AF with a Néel temperature of 160\,K.

Using a Sharrock-type expression for the dependence of the switching field on the effective island moment and the energy barrier for 180$^\circ$ Néel vector rotation, quantitative agreement with the experimental data is obtained for a tiny imbalance in top and bottom plane Cr atom moments. According to our DFT calculations the origin of the imbalance in magnetic moments is a charge transfer from the graphene substrate primarily to the bottom Cr$_\mathrm{bottom}$ atom plane of Cr$_2$S$_3$-2D.  

Our research implies that for a fixed external magnetic field the Néel vector in Cr$_2$S$_3$-2D can be controlled by gates. Being air stable and transferable, only bound by van der Waals forces to Gr, Cr$_2$S$_3$-2D possesses the essential properties for electrical switching, paving the way for future device-oriented experiments.


\clearpage
\section{Experimental Section}

\subsection{Sample preparation}

\noindent
Cr$_2$S$_3$-2D on Gr/Ir(110) is grown in Cologne in an UHV chamber (base pressure = $1 \times 10^{-10}$\,mbar), enabling UHV transfer into the SP-STM system. To grow Gr on Ir(110), it is first cleaned and prepared in its unreconstructed state using cycles of 1\,keV Ar$^+$ sputtering, flash annealing to 1510\,K, and subsequent cooling to 400\,K in $1\times10^{-7}$\,mbar oxygen pressure. Afterward, Ir(110) is heated to 1510\,K and then exposed to $3\times10^{-7}$\,mbar ethylene for 240\,s resulting in single crystal Gr on Ir(110) \cite{Kraus2022}. To grow single-layer Cr$_2$S$_3$-2D, Gr/Ir(110) is exposed to a Cr flux of $5 \times 10^{16}$\,atoms m$^{-2}$ s$^{-1}$ from an e-beam evaporator for 300\,s in a sulfur background pressure of $5\times10^{-8}$\,mbar measured by a distant ion gauge.  The S pressure is obtained by decomposing pyrite (FeS$_2$) in a Knudsen cell. After growth, the sample is subsequently annealed at 850\,K for 300\,s in the presence of a S background pressure of $5\times10^{-8}$\,mbar \cite{safeer2025}.

\noindent
\subsection{STM and SP-STM measurements}

\noindent
STM measurements are performed at 4.2\,K. Constant-current topographies and $\mathrm{d}I/\mathrm{d}V$ maps were recorded with sample bias $V_\mathrm{b}$ and tunneling current $I_\mathrm{t}$, as detailed in the respective figure captions. The lock-in technique with a modulation voltage of 20\,mV at 667\,Hz is used to record the $\mathrm{d}I/\mathrm{d}V$ maps. The average $\mathrm{d}I/\mathrm{d}V$ value from the $\mathrm{d}I/\mathrm{d}V$ maps is extracted by using Gwyddion software~\cite{Necas2012}. Spin-polarized STM tips are obtained by dipping a clean tungsten tip into a Cr$_2$S$_3$-2D island, followed by subsequent bias pulsing on Gr/Ir(110) areas. The superparamagnetic tips were calibrated on a bilayer Fe nanoisland on Cu(111), as exemplified in Figure~S1, Supplementary Information, where it provides the expected helical magnetic order in the Fe island \cite{Phark2014}.

\noindent
\subsection{XMCD measurements}

\noindent
XMCD measurements were performed at the BOREAS beamline at ALBA Synchrotron with the collaboration of ALBA staff~\cite{Barla2016}. The XMCD measurements were performed in total electron yield (TEY) mode measuring the sample drain current to ground. The signal was normalized to the incoming photon flux measured as the drain current signal on a freshly-evaporated gold mesh placed between the last optical element and the sample. The TEY signal surface sensitivity is determined by the secondary electron escape depth and is of the order of several nanometers in the Cr L$_{2,3}$ energy range. The samples were transported from Cologne to Barcelona, either stored in a UHV suitcase ($5 \times 10^{-10}$\,mbar) or an Ar-filled container. In both cases, the samples were mounted to XMCD sample holders in an Ar glove box directly connected to the XMCD UHV chamber. Finally, prior to XMCD measurements, the samples were annealed to 600\,K in $5\times10^{-8}$\,mbar S for 300\,s in a UHV preparation chamber directly connected to the XMCD facility. The XMCD results obtained did not depend on the method of sample transport. 

\noindent
\subsection{DFT calculations}

\noindent
The energetics and magnetic properties of all considered systems were investigated using spin-polarized density functional theory (DFT) as implemented in the VASP code \cite{vasp1,vasp2}.
All the calculations were performed using the Perdew-Burke-Ernzerhof exchange-correlation functional \cite{perdew_1996}. A plane-wave cut-off energy of 500 eV and force tolerance of 0.01 eV/Å was set for geometry optimisation. Van der Waals interactions were included using the DFT-D2 method proposed by Grimme \cite{DFT-D2}. A vacuum space of approximately 30 Å was introduced along the non-periodic direction to avoid interactions between periodic images. The Cr$_2$S$_3$-2D on graphene heterostructure was constructed by using an interface consisting of (5 × 5) unit cells of Cr$_2$S$_3$-2D and (7 × 7) unit cells of a graphene corresponding to a lattice mismatch of only 0.23\%. The Brillouin zone of the supercells was sampled using 4 x 4 k-points mesh. Charge transfer from graphene to Cr$_2$S$_3$-2D was calculated as follows:
\begin{equation}
\Delta \rho (z) = \rho_{\mathrm{hetero}}(z) 
- [\rho_{\mathrm{graphene}}(z) 
+ \rho_{\mathrm{Cr_2S_3}}(z)],
\end{equation}
where $\rho_{\mathrm{hetero}}$ is the charge density of the heterostructure, 
$\rho_{\mathrm{graphene}}$ represents the charge density of isolated graphene, 
and $\rho_{\mathrm{Cr_2S_3}}$ denotes the charge density of Cr$_2$S$_3$-2D. 
The charge density difference was visualized using the VESTA package~\cite{momma2008}.

\section*{Supporting Information}
\noindent
Supporting Information is available from the Wiley Online Library or from
the author.

\section*{Acknowledgements}
\noindent
Funding from the Deutsche Forschungsgemeinschaft (DFG) through CRC 1238 (project number 277146847, subprojects A01 and B06) is acknowledged. J.F. and W.J acknowledge financial support by the DFG priority programs SPP 2137 (FI 2624/1-1, project no. 462692705) and SPP2244 (project no. 535290457), respectively. A.V.K. acknowledges funding from the German Research Foundation (DFG), project KR 4866/9-1 and the collaborative research center “Chemistry of Synthetic 2D Materials” CRC-1415-417590517. A.B.-P. acknowledges support from the Generalitat Valenciana (grant CIDEGENT/2021/005) and from the Spanish MCIU (grant PID2023-149494NB- C31). The XMCD experiments were performed at the ALBA Synchrotron thanks to the official proposal (No. ID2024098828). Generous CPU time grants from the Paderborn Center for Parallel Computing (PC2, Noctua 2 cluster, hpc-prf-def2dhet) and Gauss Centre for Supercomputing eV (www.gauss-centre.eu) are greatly appreciated.

\section*{Conflict of Interest}
\noindent
The authors declare no conflict of interest.

\section*{Data Availability Statement}
\noindent
The data that support the findings of this study are available from the
corresponding author upon reasonable request.

\bibliography{Ref}

\end{document}


\DeclareGraphicsExtensions{.pdf}
\title{Supplementary Information:\\ Atomic-Scale Detection of Néel Vector Switching in the Single-Layer A-type Antiferromagnet Cr$_2$S$_3$-2D}

\author{Affan Safeer}
\email{safeer@ph2.uni-koeln.de}
\affiliation{II. Physikalisches Institut, Universit\"{a}t zu K\"{o}ln, Z\"{u}lpicher Str. 77, 50937 Cologne, Germany \looseness=-1}
\author{Calisa Dias}
\affiliation{Institute of Molecular Science, University of Valencia, Catedratico Jose Beltrán 2, 46980 Paterna, Spain}
\author{Mahdi Ghorbani-Asl}
\affiliation{Institute of Ion Beam Physics and Materials Research, Helmholtz-Zentrum Dresden-Rossendorf, 01328 Dresden, Germany
\looseness=-1}
\author{Abdallah Karaka}
\affiliation{II. Physikalisches Institut, Universit\"{a}t zu K\"{o}ln, Z\"{u}lpicher Str. 77, 50937 Cologne, Germany \looseness=-1}
\author{Pradyumna Bawankule}
\affiliation{Institute of Molecular Science, University of Valencia, Catedratico Jose Beltrán 2, 46980 Paterna, Spain}
\author{Weibin Li}
\affiliation{ALBA Synchrotron Light Source, Cerdanyola del Vallès, 08290 Barcelona, Spain
\looseness=-1}
\author{Pierluigi Gargiani 
\looseness=-1}
\affiliation{ALBA Synchrotron Light Source, Cerdanyola del Vallès, 08290 Barcelona, Spain
\looseness=-1}
\author{Wouter Jolie}
\affiliation{II. Physikalisches Institut, Universit\"{a}t zu K\"{o}ln, Z\"{u}lpicher Str. 77, 50937 Cologne, Germany \looseness=-1}
\author{Arkady V. Krasheninnikov}
\affiliation{Institute of Ion Beam Physics and Materials Research, Helmholtz-Zentrum Dresden-Rossendorf, 01328 Dresden, Germany
\looseness=-1}
\author{Amilcar Bedoya-Pinto}
\affiliation{Institute of Molecular Science, University of Valencia, Catedratico Jose Beltrán 2, 46980 Paterna, Spain}
\author{Thomas Michely}
\affiliation{II. Physikalisches Institut, Universit\"{a}t zu K\"{o}ln, Z\"{u}lpicher Str. 77, 50937 Cologne, Germany \looseness=-1}
\author{Jeison Fischer}
\email{jfischer@ph2.uni-koeln.de}
\affiliation{II. Physikalisches Institut, Universit\"{a}t zu K\"{o}ln, Z\"{u}lpicher Str. 77, 50937 Cologne, Germany \looseness=-1}


\maketitle
\clearpage
\tableofcontents

\clearpage
\subsection*{Supplementary Note 1. Spin-polarized STM tip characterization on an Fe island on Cu(111).}
\begin{figure}[hbt!]
\includegraphics[width=\textwidth]{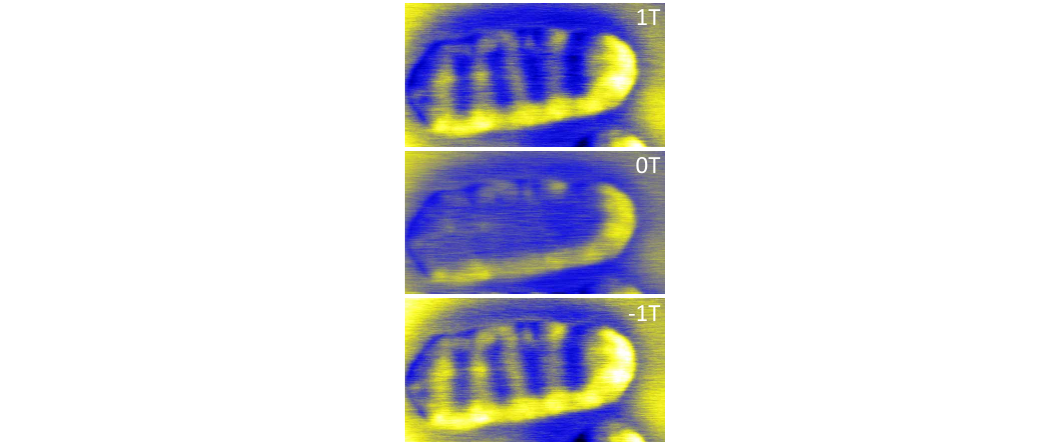}
\caption{Constant height $\mathrm{d}I/\mathrm{d}V$ map ($V_\mathrm{b} = -400$\,mV, $I_\mathrm{t} = 3$\,nA) of an Fe island on Cu(111) acquired at the magnetic fields indicated. Image sizes: 15\,nm $\times$ 7.5 \,nm. 
}
  \label{Sfig_tip_chr} 
\end{figure}

\clearpage
\subsection*{Supplementary Note 2. Spin-polarized $\mathrm{d}I/\mathrm{d}V$ spectra of single-layer Cr$_2$S$_3$-2D.}
\begin{figure}[hbt!]
\includegraphics[width=\textwidth]{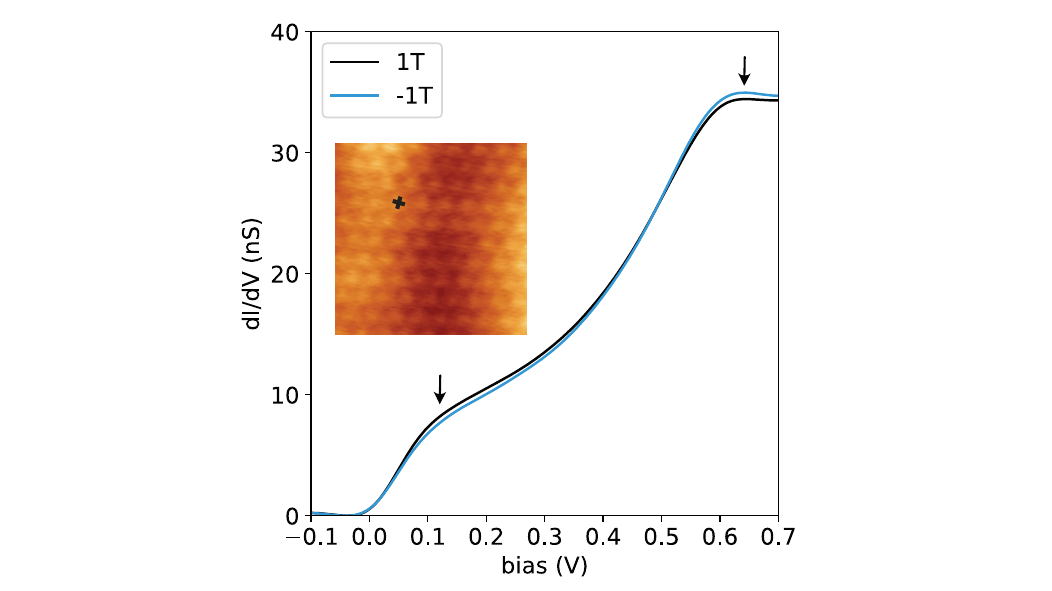}
\caption{Constant height spin polarized $\mathrm{d}I/\mathrm{d}V$ point spectra of Cr$_2$S$_3$-2D, taken at $1$~T and $-1$~T. The inset shows the atomically resolved STM image ($V_\mathrm{b} = 100$\,mV, $I_\mathrm{t} = 1$\,nA) of Cr$_2$S$_3$-2D, with a black cross indicating the location where spectra are acquired. The black arrows point to $\mathrm{d}I/\mathrm{d}V(V)$ locations with finite spin polarization. The spectra are acquired by stabilizing at $V_\mathrm{b} = 700$\,mV with setpoint current $I_\mathrm{t} = 1$\,nA. STM image size = 4\,nm $\times$ 4\,nm. 
} 
\label{Sfig_didV} 
\end{figure}

\clearpage
\subsection*{Supplementary Note 3. Average $\mathrm{d}I/\mathrm{d}V$ signal as a function of in-plane and out-of-plane magnetic field.}
\begin{figure}[hbt!]
\includegraphics[width=\textwidth]{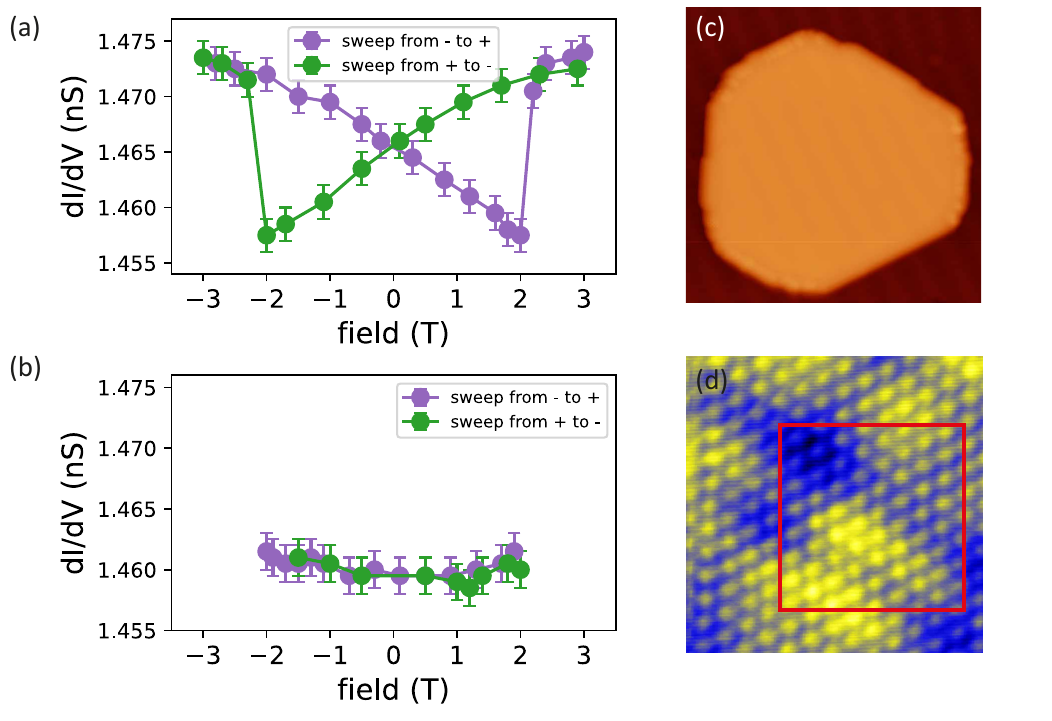}
\caption{Average $\mathrm{d}I/\mathrm{d}V$ plots as a function of (a) out-of-plane and (b) in-plane applied magnetic field. (c) STM topograph of the Cr$_2$S$_3$-2D island (size $\approx480$\,nm$^2$) on which data is acquired. (d) Constant current $\mathrm{d}I/\mathrm{d}V$ map ($V_\mathrm{b} = 100$\,mV, $I_\mathrm{t} = 1$\,nA) of the island shown in (a). The red box in (d) indicates the area used for averaging the $\mathrm{d}I/\mathrm{d}V$ signal at different magnetic fields. Image sizes: (a) 32\,nm $\times$ 32\,nm and (b) 4\,nm $\times$ 4\,nm. 
} 
  \label{Sfig_temp STM} 
\end{figure}

\clearpage
\subsection*{Supplementary Note 4. XMCD of single-layer Cr$_2$S$_3$-2D on Gr/Ir(110) for grazing incidence.}
\begin{figure}[!ht]
\centering
\includegraphics[width=\textwidth]{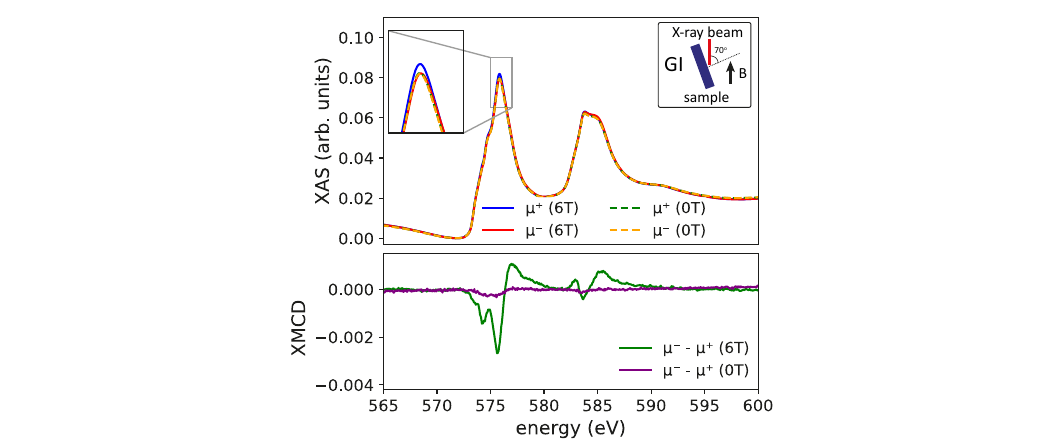}
\caption{Cr L$_{2,3}$ edges XAS and XMCD spectra of single-layer Cr$_2$S$_3$-2D on Gr/Ir(110) acquired in total electron yield mode with grazing incidence at 3.5\,K under 6\,T and 0\,T magnetic field. Insets on the left and right show the zoom of the Cr L$_3$ edge and the measurement schematics, respectively. The XMCD spectrum is obtained by subtracting $\mu$$^-$ from $\mu$$^+$ XAS.
}
\label{SFig_XMCD_GI}
\end{figure}

\clearpage
\subsection*{Supplementary Note 5. Temperature-dependent XMCD of Cr$_2$S$_3$-2D.}
\begin{figure}[hbt!]
\includegraphics[width=\textwidth]{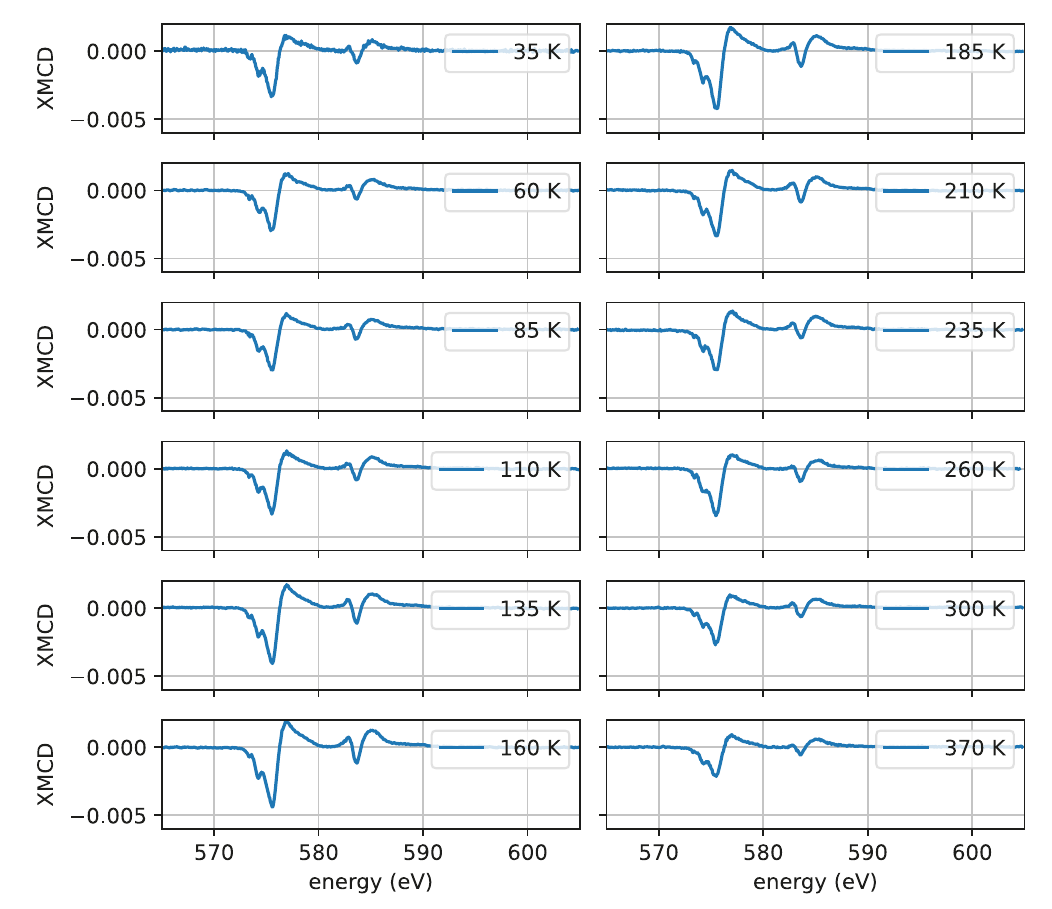}
\caption{XMCD spectra of single-layer Cr$_2$S$_3$-2D as a function of temperature, acquired with the NI X-ray beam at 6\,T magnetic field.   
} 
  \label{Sfig_temp_XMCD} 
\end{figure}

\clearpage
\subsection*{Supplementary Note 6. Temperature-dependent XLD of Cr$_2$S$_3$-2D.}
\begin{figure}[hbt!]
\includegraphics[width=0.9\textwidth]{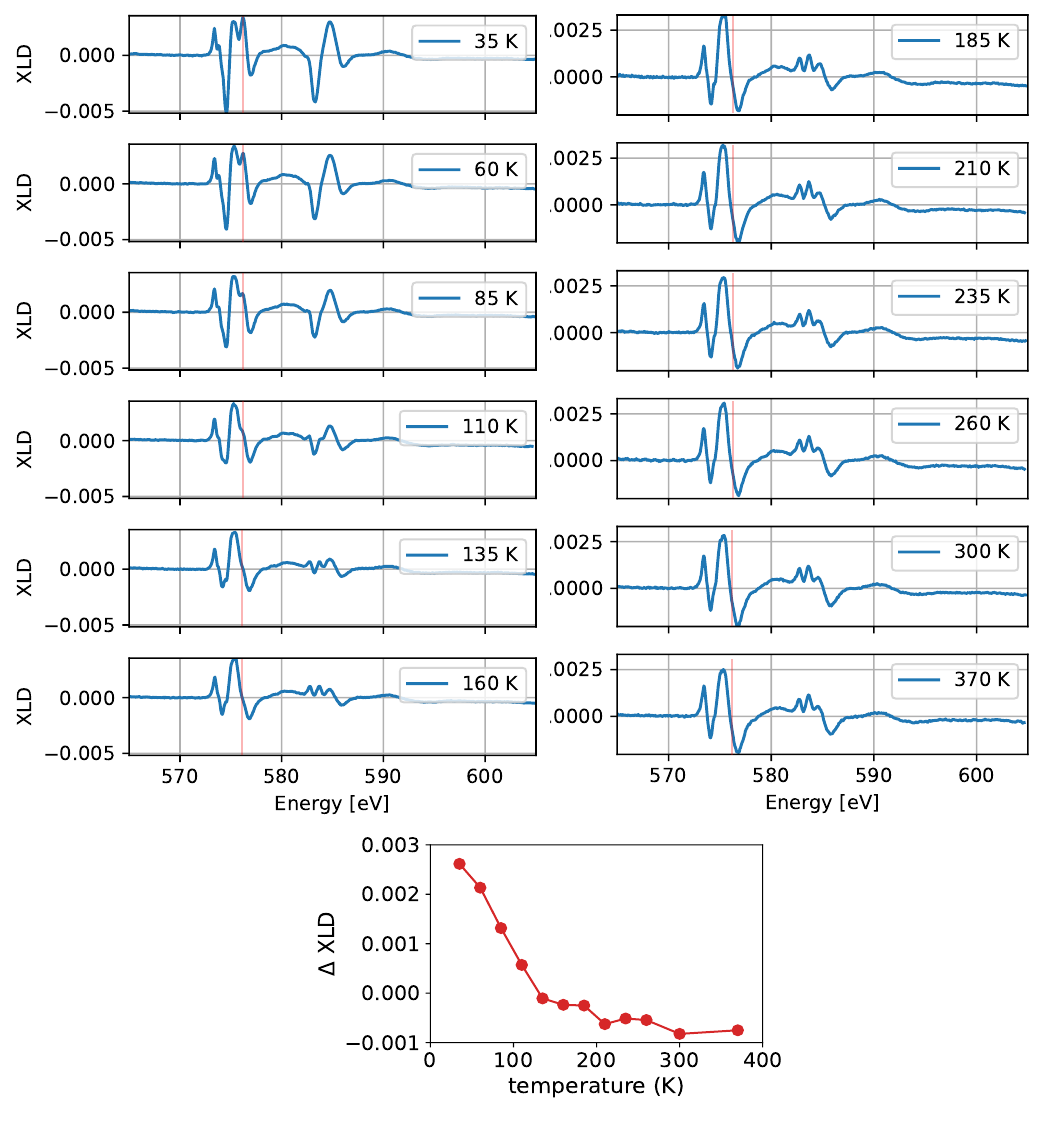}
\caption{XLD spectra of single-layer Cr$_2$S$_3$-2D as a function of temperature, acquired with the GI X-ray beam at 6\,T magnetic field. The XLD signal at 576\,eV (L$_3$ edge) is maximal at the lowest temperature and gradually decreases as the temperature approaches the Néel temperature ($T_\mathrm{N}$), becoming constant (vanishing) above. The XLD signal decrease below $T_\mathrm{N}$ is assigned to the alignment of the the Néel vector to the in-plane magnetic field in GI orientation.
} 
  \label{Sfig_temp_XLD} 
\end{figure}

\clearpage
\subsection*{Supplementary Note 7. Temperature-dependent SP-STM of  Cr$_2$S$_3$-2D.}

\begin{figure}[hbt!]
\includegraphics[width=\textwidth]{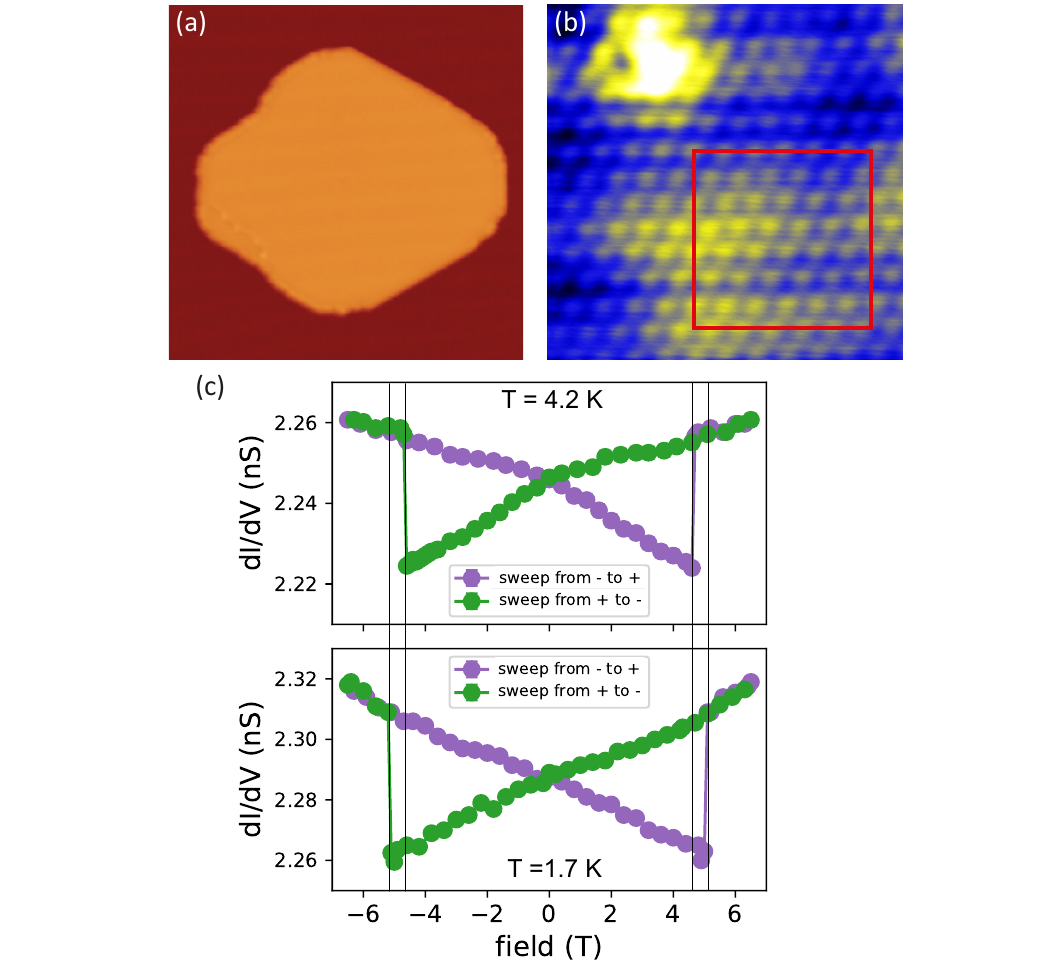}
\caption{(a) STM overview image ($V_\mathrm{b} = 1$\,V, $I_\mathrm{t} = 50$\,pA) of a Cr$_2$S$_3$-2D island on Gr/Ir(110) substrate. Constant current $\mathrm{d}I/\mathrm{d}V$ map ($V_\mathrm{b} = 100$\,mV, $I_\mathrm{t} = 1$\,nA) recorded on the Cr$_2$S$_3$-2D island shown in (a). (c) Temperature-dependent average $\mathrm{d}I/\mathrm{d}V$ value as a function of magnetic field. The average $\mathrm{d}I/\mathrm{d}V$ values are extracted from $\mathrm{d}I/\mathrm{d}V$ maps recorded at magnetic fields between $\pm$6.5\,T. The red box in (b) indicates the area used for averaging the $\mathrm{d}I/\mathrm{d}V$ signal. Image sizes: (a) 45\,nm $\times$ 45\,nm and(b) 4\,nm $\times$ 4\,nm.
}
  \label{Sfig_temp_SPSTM} 
\end{figure}

\clearpage
\subsection*{Supplementary Note 8. Energy barrier analysis with uncompensated moments arising from edges}

\begin{figure}[!ht]
\centering
\includegraphics[width=0.8\textwidth]{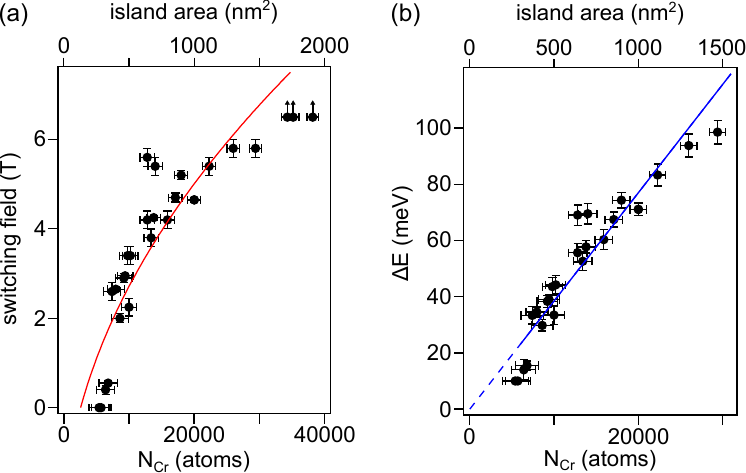}
\caption{(a) Switching field $\mu_0 H_{\mathrm{sw}}$ and (b) energy barrier ($\Delta E$) of islands as a function of island size. $\Delta E$ is extracted from $\mu_0 H_{\mathrm{sw}}$ in (a) by using equation~1 (main text). Green and blue lines in (a) and (b) are fits based on equation~1 (main text) assuming the uncompensated magnetic moment to stem from island edge Cr atoms. To compute the number of Cr atoms at the edge, we assume a circular perimeter, giving $N_\mathrm{Cr}^\mathrm{edge}=\sqrt{\sqrt{3} \pi N_\mathrm{Cr}^\mathrm{tot}}$. Consequently, the total uncompensated moment can be approximated as $\mu_u=N_\mathrm{Cr}^\mathrm{edge}\mu_B$.}
\label{fig_DeltaE_edges}
\end{figure}

\clearpage

\subsection*{Supplementary Note 9. DFT supercell Cr$_2$S$_3$-2D / Gr and comparison to freestanding Cr$_2$S$_3$-2D}

\begin{figure}[!ht]
\centering
\includegraphics[width=1\textwidth]{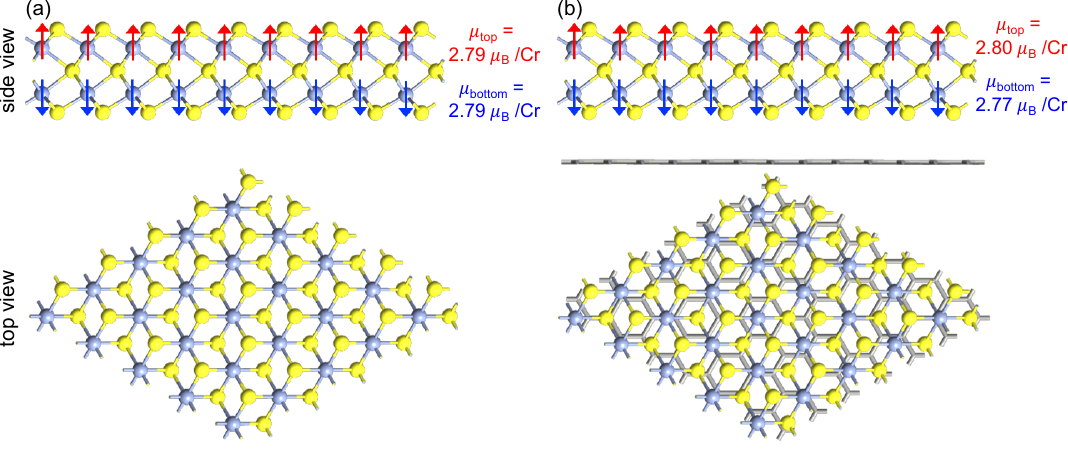}
\caption{Side and top view ball models of the Cr$_2$S$_3$-2D freestanding (a) and supercell composed of $5\times5$ Cr$_2$S$_3$-2D layer on a $7\times7$ graphene (b). In this configuration the two layers are nearly commensurate, with strain applied to Gr of 0.23\%. The DFT calculated magnetic moments of the top Cr plane $\mu_\mathrm{top}$ and bottom Cr plane $\mu_\mathrm{bottom}$ are given beside the side view ball model for each case. The values are equal for freestanding and differ by 0.03 $\mu_\mathrm{B}$/Cr for the Cr$_2$S$_3$-2D/Gr.
}
\label{fig_dft}
\end{figure}

\clearpage
\subsection*{Supplementary Note 10. Dependence of switching field on tunneling current}

\begin{figure}[!ht]
\centering
\includegraphics[width=\textwidth]{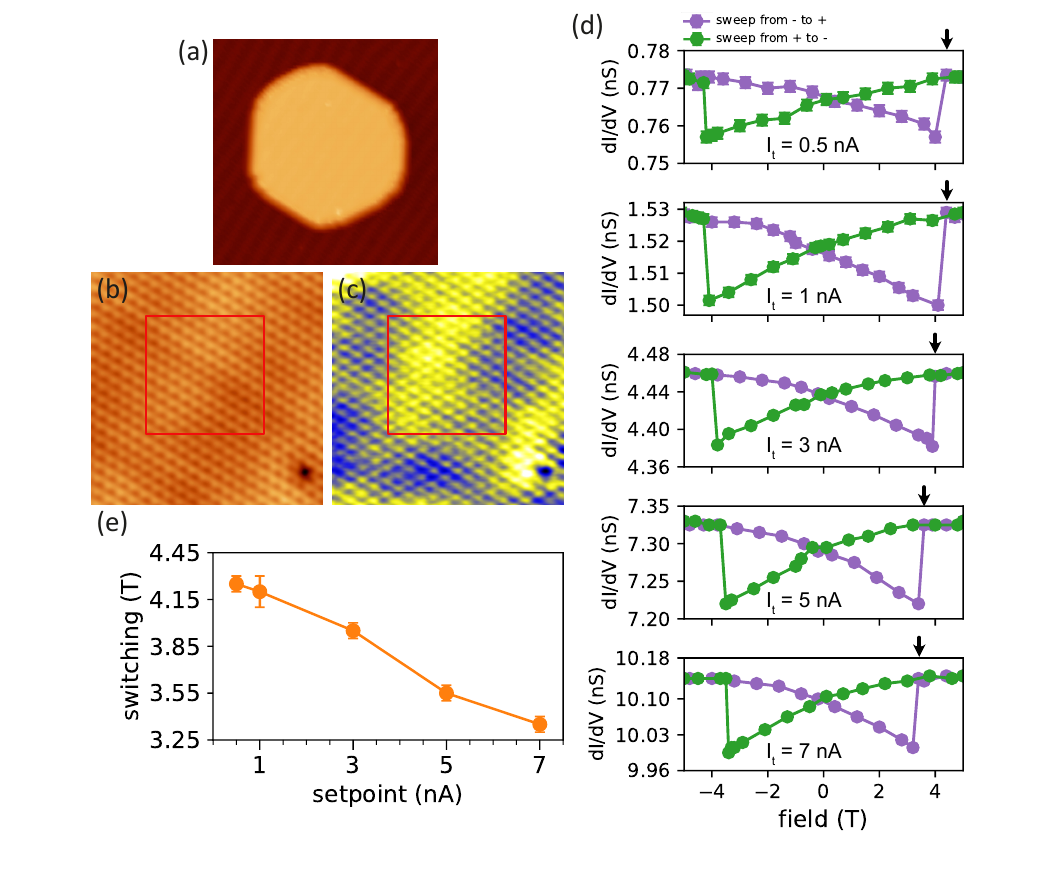}
\caption{(a) STM overview image ($V_\mathrm{b} = 1$\,V, $I_\mathrm{t} = 50$\,pA) of the Cr$_2$S$_3$-2D island on Gr/Ir(110) substrate. (b) Atomically resolved SP-STM topograph ($V_\mathrm{b} = 100$\,mV, $I_\mathrm{t} = 1$\,nA) of Cr$_2$S$_3$-2D recorded with an out-of-plane field of $B = 1$\,T. (c) $\mathrm{d}I/\mathrm{d}V$ map of Cr$_2$S$_3$-2D recorded simultaneously with (b). (d) Plot of average $\mathrm{d}I/\mathrm{d}V$ value as a function of magnetic field at $V_\mathrm{b} = 100$\,mV and $I_\mathrm{t}$ as written. Each data point in the plots is taken by averaging the $\mathrm{d}I/\mathrm{d}V$ values from the $\mathrm{d}I/\mathrm{d}V$ maps taken at magnetic fields sweeping between $\pm$6.5\,T. The red box in (c) shows the area which is used for averaging the $\mathrm{d}I/\mathrm{d}V$. (e) Switching field as a function of tunneling current. 
The red box in (b) indicates the area used for averaging the $\mathrm{d}I/\mathrm{d}V$ signal. Image sizes: (a) 50\,nm $\times$ 50\,nm and (b,c) 6.5\,nm $\times$ 6.5\,nm. }
\label{fig_DeltaE}
\end{figure}